\documentclass[letterpaper, 12pt]{article}
\usepackage{graphicx,amssymb,multicol}
\usepackage[usenames]{color}
\usepackage[small,compact,nonindentfirst]{titlesec}
\usepackage[square,numbers,sort&compress]{natbib}
\usepackage{wrapfig}
\definecolor{violet}{rgb}{0.6,0.0,0.3}
\usepackage[%
pdftitle={Exploring Dark Energy with Next-Generation Photo-z Surveys},%
pdfauthor={Zhan et al.},%
bookmarksnumbered=true,%
linkcolor=violet,citecolor=blue,urlcolor=blue,colorlinks=true]{hyperref}

\newcommand{\aj}{{AJ}}
\newcommand{\apj}{{ApJ}}

\newcommand{\apjl}{{ApJ}}
\newcommand{\mnras}{{MNRAS}}
\newcommand{\prd}{{Phys. Rev. {\rm D}}}

\newcommand{\aap}{{A\&A}}

\newcommand{\phz}{photo-\emph{z}}
\newcommand{\Phz}{Photo-\emph{z}}
\newcommand{\wa}{w_{\rm a}}
\newcommand{\Ok}{\Omega_{\rm k}}
\newcommand{\Om}{\Omega_{\rm m}}

\newcommand{\remove}[1]{}

\titlelabel{}
\titleformat*{\section}{\vspace*{-2.5ex}\center\large\bf}
\titleformat*{\subsection}{\bf}
\titlespacing{\section}{0pt}{*1}{0.6ex plus 0.2ex minus 0.2ex}
\titlespacing{\subsection}{0pt}{0.8ex plus 0.2ex minus 0.2ex}
{0.5ex plus 0.2ex minus 0.2ex}

\newcommand{\captionfonts}{\linespread{1}\footnotesize\bf}

\makeatletter  % Allow the use of @ in command names
\long\def\@makecaption#1#2{%
  \vskip\abovecaptionskip
  \sbox\@tempboxa{{\captionfonts #1: #2}}%
  \ifdim \wd\@tempboxa >\hsize
    {\captionfonts #1: #2\par}
  \else
    \hbox to\hsize{\hfil\box\@tempboxa\hfil}%
  \fi
  \vskip\belowcaptionskip}
\makeatother   % Cancel the effect of \makeatletter

\begin{document}
\pagestyle{empty}

\begin{center}
{\Large \bf Exploring Dark Energy with Next-Generation \\[0.8ex]
Photometric Redshift Surveys}

\vspace{4ex}

Hu Zhan (UC Davis), Andreas Albrecht (UC Davis),
Asantha Cooray (UC Irvine), \\
Salman Habib (LANL), Alan Heavens (U. Edinburgh), 
Katrin Heitmann (LANL), Bhuvnesh Jain (UPenn), 
Myungkook J.~Jee (UC Davis), Lloyd Knox (UC Davis), \\
Rachel Mandelbaum (IAS), Jeff Newman (U. Pitt), 
Samuel Schmidt (UC Davis), \\
Ryan Scranton (UC Davis),  Michael Strauss (Princeton),
Tony Tyson (UC Davis), \\ Licia Verde (UAB \& Princeton), 
David Wittman (UC Davis), and \\ Michael Wood-Vasey (U. Pitt)

\vspace{0.4in}

\begin{minipage}{6.2in} 
Contact: Hu Zhan, Department of Physics, University of 
California, Davis, CA 95616\\
\phantom{Contact:} (530) 754-6928, 
\href{mailto:hzhan@ucdavis.edu}{hzhan@ucdavis.edu}
\end{minipage}

\end{center}

\vspace{1.5in}
\begin{center}
\includegraphics[width=6in]{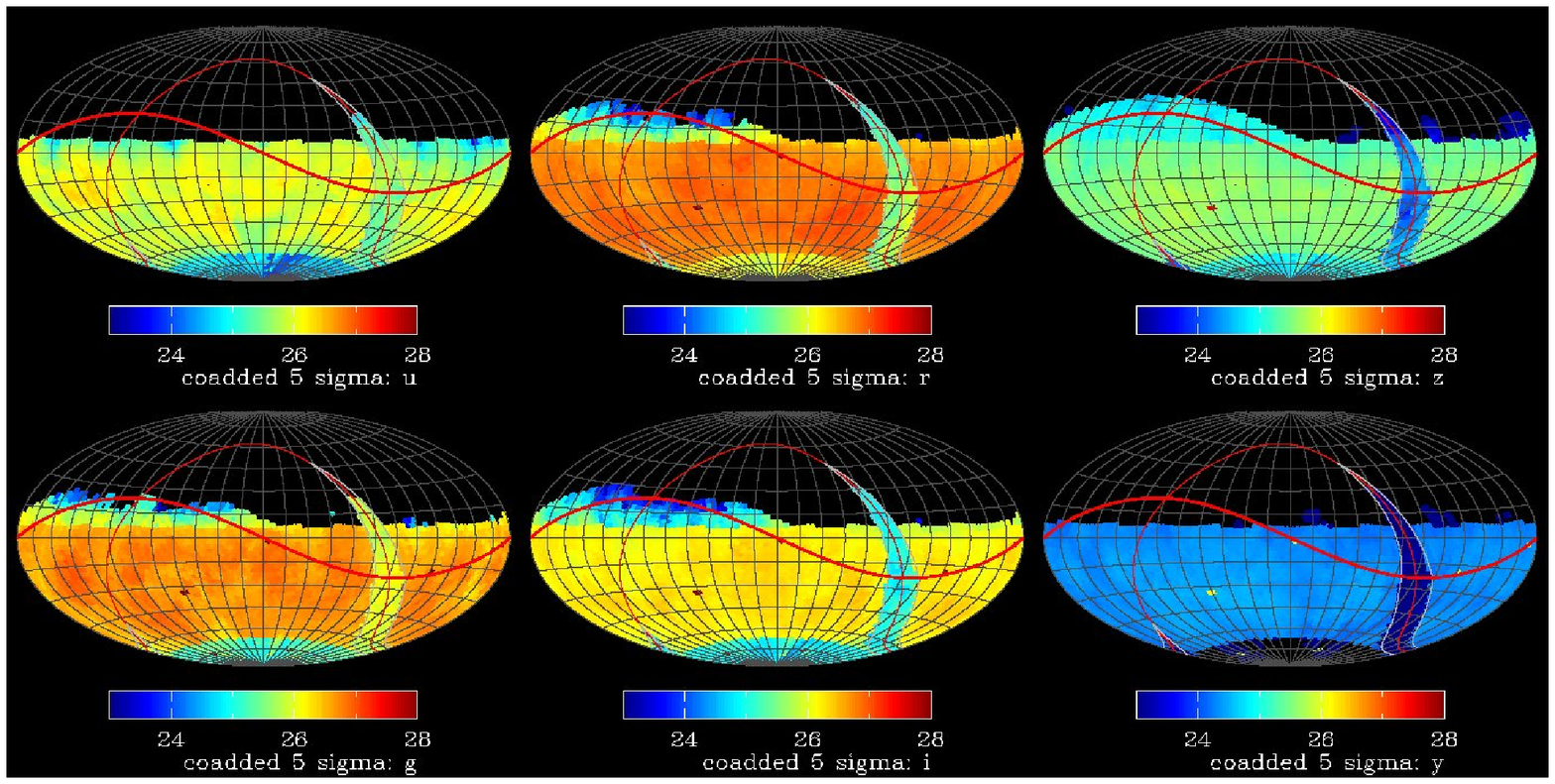}
\end{center}

%\tableofcontents

\newpage
\pagestyle{plain}
\setcounter{page}{1}

\phantom{x}
\vspace{-3.6ex}

\section{Summary}

The coming decade will be an exciting period for dark energy research, 
during which astronomers will address the question of {\bf what drives 
the accelerated cosmic expansion} as first revealed by type Ia supernova 
(SN) distances \citep{riess98perlmutter99}, and confirmed by later 
observations.

The mystery of dark energy poses a challenge of such magnitude that, as 
stated by the Dark Energy Task Force (DETF), ``\emph{nothing short of a 
revolution in our understanding of fundamental physics will be required 
to achieve a full understanding of the cosmic acceleration}'' 
\citep{albrecht06}. The lack of multiple complementary 
precision observations is a major 
obstacle in developing lines of attack for dark energy theory. This lack 
is precisely what next-generation surveys will address via the powerful 
techniques of weak lensing (WL) and baryon acoustic oscillations (BAO) 
-- galaxy correlations more generally -- 
in addition to SNe, cluster counts, and other probes of geometry and 
 growth of structure. Because of their unprecedented statistical power, these 
 surveys demand an accurate understanding of 
the observables and tight control of systematics.

This white paper highlights the opportunities, approaches, prospects, 
and challenges relevant to dark energy studies with wide-deep
multiwavelength
photometric redshift (\phz{}) surveys. Quantitative predictions are 
presented for a 20000 deg$^2$ ground-based 6-band (\emph{ugrizy}) 
survey with $5\sigma$ depth of $r\sim 27.5$ 
\citep{ivezic08}, i.e., a Stage 4 survey as defined by the DETF.

\section{Exploring Dark Energy with Multiple Probes
\label{sec:mprb}}

Cross-checks and confirmations by multiple lines of evidence are 
extremely important in cosmology. This is especially true for dark
energy investigations, as the properties of dark energy can only be inferred
from often subtle effects. In fact, most
future surveys are designed to enable multiple techniques 
using a single, uniform data set and allow for cross-correlating with 
other types of observations, so that they can form interlocking 
cross-checks on the accelerated expansion and maximize the science
output. For example, WL has a stringent requirement on image quality, 
so a wide-area WL survey is readily suitable for angular BAO, cluster 
counting, strong lensing, and, with a proper design of survey 
programs, SNe as well.

These techniques probe the cosmic acceleration through
its effect on the growth of structure and/or geometry of the 
Universe. They have different parameter degeneracies and vary in 
sensitivity over redshift.  By comparing results of the same quantity 
(such as distance)
from multiple probes, one can detect and possibly rectify 
unexpected systematics of each probe. By combining them,
one can break individual degeneracies and achieve stronger 
constraints on dark energy properties. Moreover, a multi-probe 
approach allows one to test the consistency of dark energy models
and constrain (or explore) new physics \citep[e.g.,][]{wanghui07}.

\begin{figure}
\centering
\includegraphics[width=5.7in]{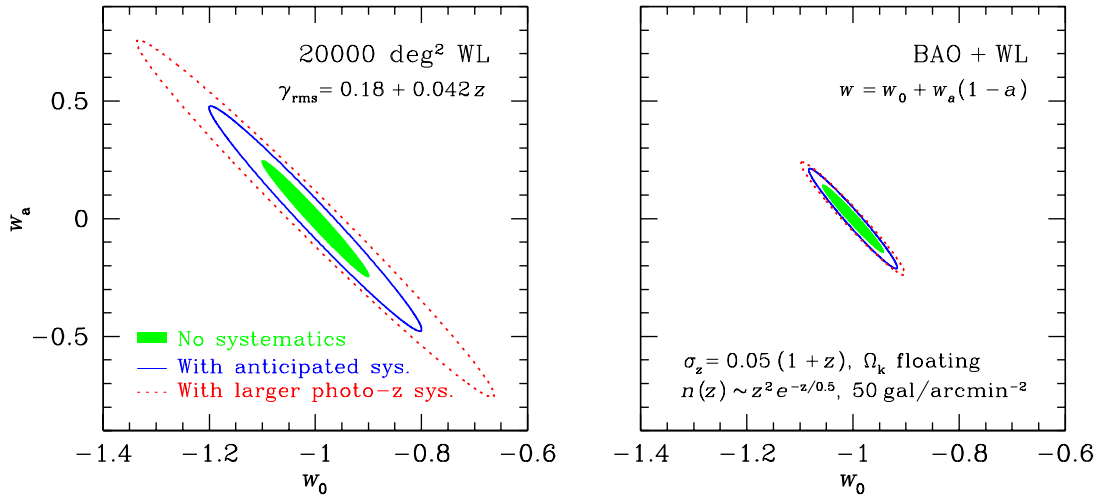}
\caption{Complementarity between WL and BAO.
The dramatic improvement of the BAO+WL results (right)
over the WL-alone results (left) is due to the cross-calibration 
of galaxy bias and \phz{} uncertainties 
and is independent of the dark energy EOS parametrization.
The results are marginalized over 9 other
cosmological parameters and over 140 parameters
for the galaxy bias, \phz{} bias, \phz{} rms, and 
additive and multiplicative errors of the power spectra
\citep{huterer06,ma06,zhan06,zhan09}.
\label{fig:wsys}}
\end{figure}

Correlations can be significant between some probes. If they are 
not properly accounted for, the combined constraining power will be 
over-estimated. However, correlations can provide 
useful information as well. In the case of BAO and WL, 
a joint analysis of the shear and 
galaxy overdensities for the same set of galaxies involves 
galaxy--galaxy, galaxy--shear, and shear--shear correlations, which
enable some calibration of systematics that would otherwise adversely 
impact each probe \citep{zhan06}. Figure~\ref{fig:wsys} 
demonstrates that while the WL constraints on the dark energy equation 
of state (EOS, $w=p/\rho$) parameters, $w_0$ and $\wa$, as defined by 
$w=w_0+\wa(1-a)$, are sensitive to systematic 
uncertainties in the \phz{} error distribution, the joint 
BAO and WL results remain fairly immune to these systematics. 

The $w_0$--$\wa$ parametrization in Fig.~\ref{fig:wsys} 
 does not capture the complexity of all dark energy models. 
It also significantly underestimates the full capabilities of Stage 4
surveys \citep{albrecht07}. More generally,
one may allow the EOS to vary independently at different redshifts
and let the data determine the EOS eigenmodes and their errors, 
which can then be used to constrain dark energy models.
Stage 4 surveys can measure at least 5 EOS eigenmodes with errors
better that 10\% each, and may completely 
eliminate some quintessence models \citep[e.g.,][]{barnard08}.

We emphasize that the projection in Fig.~\ref{fig:wsys} assumes
not only progress in observation and theory  but also a facility
designed and engineered to deliver superb image quality with high 
throughput. Billions of galaxies are required. We also 
note that a very-deep-and-wide survey has the unique capability of 
exploring very-large-scale properties of the Universe; this topic
is discussed in detail in a separate Wide-Field Cosmology white paper
by Scranton \emph{et~al}.

\vspace{0.5ex}

\section{Dark Energy Study Prospects\label{sec:prosp}}

\subsection{Precision Measurements of Distance, Growth, and Curvature
\label{sec:dgc}}

Dark energy properties are derived from variants of the 
distance--redshift and growth--redshift relations. Different dark
energy models feature different parameters, and various 
phenomenological parametrizations may be used for the same  
quantity such as the EOS. In contrast, distance and growth 
measurements are model-independent, as long as dark energy does 
not alter the matter power spectrum directly. Hence, it is 
desirable for future surveys to provide results of the distance and 
growth of structure, so that different theoretical models can 
be easily and uniformly confronted with the data.

\begin{wrapfigure}[18]{r}{2.8in}
\centering
\includegraphics[width=2.787in]{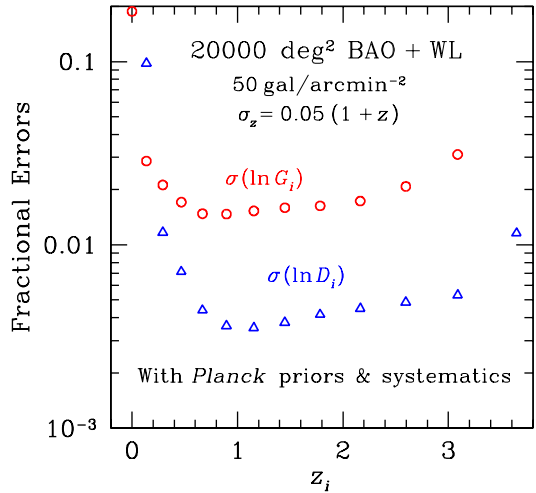}
\caption{Joint BAO and WL constraints on the comoving distance 
(open triangles) and growth factor (open circles) from the fiducial 
survey \citep{zhan09}. 
\label{fig:dges}}
\end{wrapfigure}

Figure \ref{fig:dges} demonstrates for the fiducial survey that 
joint BAO and WL can achieve 
$\sim 0.5\%$ precision on the distance and $\sim 2\%$ on the growth 
factor from $z = 0.5$ to $3$ in each interval of $\Delta z \sim 0.3$ 
\citep{zhan09}. Such measurements can test the consistency of dark 
energy or modified gravity models 
\citep[e.g.,][]{knox06a,Heavens07}.

The mean curvature of the Universe has a significant impact on dark 
energy measurements. Allowing the curvature parameter $\Ok$ to
float greatly weakens the ability of SNe to 
constrain $\wa$ 
\citep{linder05b,knox06b}. For the fiducial survey, BAO and WL can 
determine $\Ok$ to $\sim 10^{-3}$ separately and $< 10^{-3}$ jointly, 
and their results on $w_0$ and $\wa$ are not affected in practice 
by the freedom of $\Ok$ \citep{zhan06,knox06b}. 
Given its large area, the fiducial 
survey can place a tight upper limit on curvature 
fluctuations, which are expected to be small ($\sim 10^{-5}$) at 
the horizon scale in standard inflation models.

\subsection{Is Dark Energy A Cosmological Constant? \label{sec:lambda}}
\vspace{-2.8ex}

The results of current observations are consistent with a cosmological 
constant, $\Lambda$, as the cause of the cosmic acceleration. However, 
the magnitude of $\Lambda$ as observed is completely inconsistent with 
any theoretical expectation. This situation has 
driven the development of other explanations for cosmic acceleration, 
although none so far are compelling. It is imperative that observations 
be able to  distinguish between a cosmological constant and 
a dynamical origin for the dark energy. Either result may be the 
starting point for a revolution in our 
understanding of fundamental physics.

The cosmological constant, with $w = -1$, must be constant in 
space-time and admit no interaction with matter or fields. 
Therefore, detection of $w \ne -1$, time evolution of $w$, 
anisotropy or inhomogeneity, or 
nonstandard gravitational effects
will indicate non-$\Lambda$ dark energy or new physics (see below). 

\subsection{Is Dark Energy Isotropic and Homogeneous? \label{sec:iso}}

The easiest test of the isotropy of dark energy is to measure its 
properties in different directions on the sky with a 
\emph{highly uniform survey}. 
For example, one can determine the dark energy EOS in 
thousands of pixels across the sky with each containing a few 
hundred SNe.  Then one can examine the distribution, mean, and 
rms value of $w$ over all the pixels to see if dark energy appears 
isotropic in space from our vantage point. 
This approach can potentially provide a high-resolution map of the 
dark energy EOS on the sky, but for a given survey one 
has to trade resolution with the precision of the EOS in each pixel. 
A wide-deep BAO+WL survey of billions of galaxies is even more 
powerful (see the Wide-Field Cosmology white paper).

To examine the homogeneity, we can attempt to measure the clustering
of dark energy, as parameterized by the sound speed ($c_s$) of the
dark energy fluid.  For $c_s \ge 1$, dark energy will remain
smooth as the clustering scale for the fluid keeps pace with the
expanding sound horizon of the universe.  If $c_s < 1$, then
eventually the horizon will catch up to the clustering scale of dark
energy, at which point it will begin to fall into the largest
gravitational potentials.  Since our current measurements are
consistent with a very smooth dark energy \cite{Giannantonio08}, we
would expect to see the signature of clustering dark energy most
directly in the integrated Sachs-Wolfe (ISW) effect through 
CMB--galaxy correlations.  The ISW effect on CMB photons is
generated by the decay of gravitational potentials due to the
expansion of the universe exceeding the clustering rate of dark
matter.    For a flat universe, detecting an ISW signal is a strong
indicator of the existence of dark energy.  However, if $c_s < 1$,
then dark energy will begin falling into the largest potentials at
very late times, suppressing the ISW signal on those scales.  For a
$w=-0.8$ cosmology that is otherwise consistent with $\Lambda$CDM, 
our fiducial survey together with 
\emph{Planck} would be able to constrain the smoothness of dark
energy on scales of 1 Gpc to better than 10\% \cite{Hu04}.  The
sensitivity of detection is a strong function of the exact value of
$w$ and $c_s$, but this measurement offers the best possible chance at
detecting dark energy inhomogeneities.

\subsection{Is Acceleration Caused by Modified Gravity Instead?
\label{sec:mg}}

Alternative explanations for the apparent cosmic acceleration
(e.g., modified gravity or the effect of an inhomogeneous 
background metric) are being actively investigated. 
For demonstration purposes, we show here how well the fiducial 
survey can distinguish modified gravity from a dark energy that  
presumably preserves the framework of General Relativity (GR).

Like dark energy, modified gravity alters the distance--redshift and 
growth--redshift relations, so multiple dark energy probes are also 
probes of modified gravity. While it is always possible to find an EOS 
for dark energy that allows the expansion history to be accounted 
for within GR: 
$w(a) = -(1/3)d \ln[\Om^{-1}(a)-1]/d\ln a$, where $\Om(a)$ is the 
matter fraction, the growth of structure generally differs in 
different models. Thus, growth measurements 
\begin{wrapfigure}[24]{r}{2.8in}
\centering
\includegraphics[width=2.787in]{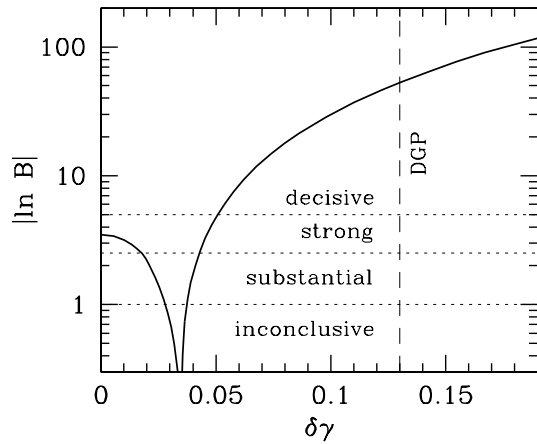}
\caption{Bayesian evidence $B$ for GR as a function of the true
deviation of the growth index from GR, 
$\delta \gamma = \gamma - 0.55$, for a Stage 4 WL survey comparable to 
our fiducial survey in combination with \emph{Planck} \cite{Heavens07}.
The larger the $B$ value, the greater the statistical power of this 
survey to distinguish the models. If modified gravity is the true
model, GR will be favored by the data to the left of the cusp
($B>1$), and increasingly disfavored to the right ($B<1$). 
The Jeffreys scale of evidence \citep{Jeffreys61} is as labeled.
Joint BAO and WL will place stronger constraints.
\label{dgamma}}
\end{wrapfigure}
are crucial 
for finding evidence for beyond-Einstein gravity (see the Wide-Field
Cosmology white paper for a discussion of other tests). 

In the convenient minimal modified gravity parametrization, the
deviation from GR is 
captured in a growth index $\gamma$ \cite{Linder05}.
In the standard GR cosmological model, $\gamma\simeq 0.55$, 
whereas in modified gravity 
theories it deviates from this value.  For a strawman example, 
the flat DGP braneworld model \citep{DGP} has $\gamma\simeq 0.68$
on scales much smaller than those where cosmological 
acceleration is apparent \citep{LinderCahn07}.

Measurements of the growth factor (e.g., Fig.~\ref{fig:dges}) 
can be used to
determine the growth index $\gamma$ and constrain modified gravity
models. In terms of model selection, one may compare a dark energy
model that has a fixed GR value for $\gamma$ with a modified gravity
model whose $\gamma$ is determined by the data and ask 
``do the data require the additional parameter and therefore 
signal the presence of new physics if the 
new physics is actually the true underlying model?'' This question may 
be answered with the Bayesian evidence, $B$, which is the ratio of 
probabilities of two or more models, given some data.

Figure~\ref{dgamma} shows how the Bayesian evidence for GR changes 
with increasing true deviation of $\gamma$ from its GR value 
for a combination of a Stage 4 WL survey (comparable to 
our fiducial survey) and \emph{Planck} \citep{Heavens07}. It is
assumed that the expansion history in the modified gravity 
model is still well described by the $w_0$--$\wa$ parametrization.
The combination of WL and \emph{Planck} could strongly distinguish 
between GR and minimally-modified gravity models whose growth index 
deviates from the GR value by as little as $\delta \gamma = 0.048$. 
Even with the WL data alone, one should be able to decisively 
distinguish GR from the DGP model at $\ln B \simeq 11.8$, or, 
in the frequentist view, $5.4\sigma$ \citep{Heavens07}.
Joint BAO and WL will place even stronger constraints.

\section{Enabling Next-Generation Dark Energy Studies \label{sec:sys}}

Unprecedented prospects will face unprecedented challenges. Three key 
issues must be faced to fully realize the science potential.

\subsection{Breakthrough Facility}

As emphasized by DETF, a next generation facility is required for this 
science mission.  Deep multiwavelength coverage of half the sky is 
required for these tests of dark energy, together with uniform good
image quality. The Large Synoptic Survey Telescope (LSST) 
\citep{ivezic08} with active optics is designed for this application.  
Variants of JDEM are synergistic, providing near-IR photometry.

\subsection{Controlling Systematics}

\Phz{} errors are one of the most critical systematics for an imaging 
survey, as redshift errors directly affect the interpretation 
of the distance--redshift and growth--redshift relations.
The effects of \phz{} errors are twofold: they randomize galaxy 
positions in the line-of-sight direction, causing a loss of 
information, and the uncertainty
in the error distribution leads to uncertain predictions of 
the observables (or biases in the analysis, if the uncertainty 
is underestimated). 

Currently, \phz{}s from ground-based observations have rms errors of 
$\sigma_z \sim 0.05(1+z)$ per galaxy for $0<z<3$. 
Future surveys will do better with deeper imaging, more precise 
photometric calibration, and larger spectroscopic training samples. 
Adding very deep near-infrared photometry, which can be obtained if 
JDEM covers the same survey area in \emph{JHK},
will reduce the \phz{} errors, particularly at $1.5 < z < 2.5$ 
\citep{abdalla08}.
A more important task is to calibrate the \phz{} error distribution 
and model it realistically in parameter estimation. 
Direct calibration with spectroscopy
is impractical for the faintest galaxies in the photometric 
catalog. Indirect methods that utilize cross-correlations
between spatially overlapping spectroscopic and photometric samples 
\citep{newman08} or those between different photometric samples 
\citep{zhan06,schneider06} do not require deep spectroscopic 
sampling and hold promise for application to future surveys.

For WL, shear measurement errors are another source of 
systematics. They are characterized by a multiplicative 
factor (or shear calibration bias), and an additive component, which is 
caused mainly by imperfect correction of the anisotropic point 
spread function. 
Current methods consistently achieve smaller than $2\%$ shear 
calibration bias \citep{massey07a}. We project that 
it can be reduced to $0.5\%$ in $\sim 5$ years. 
The additive error is correlated over small angles, which is
potentially problematic, but the impact on dark energy measurements 
will be small if its amplitude is sufficiently low 
\citep{huterer06}. Extensive ray-tracing simulations, in 
which photons travel through turbulent atmospheric layers and 
realistic optics with conservative margins of fabrication errors 
(e.g., chip tilt), show that the shear additive error
will be several orders of magnitude lower than cosmic shear on 
scales of interest for next-generation WL surveys \citep{jee09}.
This is supported by a study with Subaru observations \citep{wittman05}.

\subsection{Predicting the Observables \label{sec:obs}}

Predicting the properties of the observables given a cosmological 
model is crucial for data interpretation and statistical inference. 
Uncertainties in the predictions, if not resolved satisfactorily, 
will undermine the tremendous statistical power of 
future surveys. Computational cosmology and cross-calibration with 
different observations are indispensable and 
\emph{complementary} tools for meeting this challenge.

The key requirements for cosmological simulations are (1) to cover 
sufficiently large volumes with appropriate mass and spatial 
resolution set by the survey, (2) to include the relevant physics -- 
gravity and astrophysical processes, and (3) to return 
results with accuracies that match the survey requirements. 
The accuracy targets are  demanding, being 
$\lesssim 1\%$ around $k = 1\,\mbox{$h^{-1}$Mpc}$ 
for the matter power spectrum \citep{huterer05}. 

In the gravity-only case, the challenge will likely be met 
over the next few years since simulation accuracy, parametric 
reach, and simulation size requirements are well within the 
capabilities of petascale supercomputers \citep{heitmann08}. 
The addition of gas physics is problematic not only from 
the point of view of numerical accuracy and complexity, but also 
because of our lack of detailed knowledge about basic processes
(e.g., star formation). 
For instance, hydrodynamical simulations show that baryonic
effects may be significant on scales of interest for some WL 
surveys \citep{jing06,rudd08}, but the results differ considerably.
Nevertheless, the baryonic effects on the matter power spectrum
can be modeled by a modification to halo profiles \citep{zentner08}, 
which will be measured accurately with the same WL survey through 
galaxy and cluster lensing on scales from well 
within to beyond the virial radius \citep{2008JCAP...08..006M}.

A theoretical uncertainty for WL is the intrinsic alignment 
of galaxy shapes with local tidal fields and/or
the larger scale cosmic web. 
Low-redshift observations suggest that intrinsic alignments may 
systematically contaminate the cosmic shear power spectrum by a 
few percent at $\ell \sim 500$ for Stage 4 WL surveys 
\citep{mandelbaum06hirata07}, comparable to the statistical errors. 
Therefore, these alignments must be removed from the data.
Several promising
schemes to remove these alignments \cite{king05joachimi08} exist that,
with the addition of prior information to
be available in the next few years \cite{2007NJPh....9..444B}, allow
for reduction of the intrinsic alignment contamination without too much
loss of information.  Finally, other observations, such as 
galaxy-galaxy lensing, three-point functions, and shear B-modes, can 
reduce the effects of intrinsic alignments
\cite{heymans06semboloni08}.   

\subsection{Mining Huge, Complex Datasets}

As learned from CMB experiments, 
we need support for development of data analysis methods capable of 
tackling the challenge of extracting the influence of subtle effects:
observables affected by dark energy are also affected by 
other ``nuisance'' parameters. Depending on how the analysis is 
done, these other parameters number anywhere from tens to hundreds.  
Some of the 
nuisance parameters are cosmological, such as the density of dark matter 
today, and some are astrophysical/phenomenological; e.g., those that 
govern the \phz{} distributions.  In order to reach
conclusions about the dark energy parameters we need to run numerical 
simulations of the data and we require analysis methods that make the 
most efficient use of these simulations.  The simulations are necessary 
not only for understanding how the statistics derived from the data 
(such as the shear power spectra) depend on the parameters, 
but also for understanding the uncertainties in those statistics.

\section{Conclusion}

Exciting science opportunities are on the horizon. 
To realize these opportunities, the community needs to 
invest in facilities such as the LSST 
and in research programs that will 
improve our understanding of the systematics, our knowledge of the 
observables, and our ability to analyze the data. 
Science frontiers are often opened by unexpected discoveries. 
The LSST is designed and optimized not only for the 
known science drivers such as dark energy but also for the capacity 
to discover new science. 
We are optimistic that the challenges associated with the 
opportunities will be met in the next few years.

\vspace{-2ex}
\begin{center}
\rule{3in}{0.5pt}
\end{center}
\vspace{-6.3ex}

%\bibliographystyle{prsty}
%\bibliographystyle{apj}
%\bibliography{ref}

\end{document}